\documentclass[11pt,a4paper]{article}
\begin{document}

\title{ The stable problem of the black-hole connected region
in the Schwarzschild black hole \footnote{E-mail of Tian:
hua2007@126.com, tgh-2000@263.net}}
\author{Guihua Tian$^{1,2}$\\
1.School of Science, Beijing University \\
of Posts And Telecommunications. Beijing100876, China.\\2.Academy
of Mathematics and Systems Science,\\ Chinese Academy of
Sciences,(CAS) Beijing, 100080, P.R. China}
\date{October 18, 2004}
\maketitle

\begin{abstract}
The stability of the Schwarzschild black hole is studied. Using
the Painlev\'{e} coordinate,  our region can be defined as the
\textbf{\emph{black-hole-connected region}} ($r>2m$, see text) of
the Schwarzschild black hole or the
\textbf{\emph{white-hole-connected region}} ($r>2m$, see text) of
the Schwarzschild black hole. We study the stable problems of the
\textbf{\emph{black-hole-connected region}}. The conclusions are:
(1) in the \textbf{\emph{black-hole-connected region}}, the
initially regular perturbation fields must have real frequency or
complex frequency whose imaginary must not be greater than
$-\frac1{4m}$, so the \textbf{\emph{black-hole-connected region}}
is stable in physicist' viewpoint; (2) On the contrary, in the
mathematicians' viewpoint, the existence of the real frequencies
means that the stable problem is unsolved by the linear
perturbation method in the \textbf{\emph{black-hole-connected
region}}.

\textbf{PACC:0420-q}
\end{abstract}

Studies on the stability of the Schwarzschild black hole are of
great importance both in theoretical and cosmological
back-grounds. The Schwarzschild black hole is the only candidate
for the spherically static  vacuum space-time. It is generally
believed that it is the ultimate fate of massive star after
getting off its angular momentum. Also, Many theoretical results
heavily rely on their applications to the the Schwarzschild black
hole.

Regge and Wheeler first studied the problem, and divided the
perturbations into odd and even ones \cite{rw}. The odd one is
really the angular perturbation to the metric, while even one
 corresponds to the radial perturbation to the metric
\cite{chan}. The odd perturbation equation is the well-known
Regge-Wheeler equation.

Vishveshwara made the study further by transforming the
perturbation quantities to the Kruskal reference frame, and tried
to find the real divergence at $ r=2m $ from the spurious one
caused by the improper choice of coordinate due to the
Schwarzschild metric's ill-defined-ness at $ r=2m $ \cite{vish}.
Later, Price also studied the problem carefully \cite{pric} and
Wald studied from the mathematical background \cite{wald}. In this
paper, we give full consideration on  the perturbation fields with
complex frequency. Using the Painlev\'{e} coordinate,  our region
can be defined as the \textbf{\emph{black-hole-connected region}}
($r>2m$, see text) of the Schwarzschild black hole or the
\textbf{\emph{white-hole-connected region}} ($r>2m$, see text) of
the Schwarzschild black hole. We study the stable problems of the
two kinds of region.

In reference \cite{stew}, Stewart applied the Liapounoff theorem
to define dynamical stability of a black-hole. First, according
Stewart, the normal mode of the perturbation fields to the
Schwarzschild black-hole is the perturbation fields $\Psi$ with
time-dependence of $e^{-ik t}$ which are bounded at the boundaries
of the event horizon $r=2m$ and the infinity $r\rightarrow \infty
$. The range of permitted frequency is defined as the spectrum $S$
of the Schwarzschild black-hole. Then, for the Schwarzschild
black-hole, it could be obtained by  the Liapounoff theorem
that\cite{stew}:

(1)if $\exists k \in S$ with $\Im k >0$, the Schwarzschild
black-hole is dynamically unstable,

(2)if $\Im k <0$ for $\forall k \in S$ , and the normal modes are
complete, then, the Schwarzschild black-hole is dynamically
stable,

(3)if $\Im k \leq 0$ for $\forall k \in S$  , and there is at
least one real frequency $k \in S$, the linearized approach could
not decide the stability of the Schwarzschild black-hole.

Here, we reconsider the  stability problem of the Schwarzschild
black hole using the Painlev\'{e} coordinate metric(see
following). The conclusions are:  (1) in the
\textbf{\emph{black-hole-connected region}}, the initially regular
perturbation fields must have real frequency or complex frequency
whose imaginary must not be greater than $-\frac1{4m}$, so the
\textbf{\emph{black-hole-connected region}} is stable in usual
physicist' viewpoint; (2) On the contrary, in the mathematicians'
viewpoint \cite{stew}, the existence of the real frequencies means
that the stable problem is unsolved by the linear perturbation
method in the \textbf{\emph{black-hole-connected region}} too.

First, we give a brief introduction of the odd perturbation of the
Schwarzschild black-hole\cite{rw}. The Schwarzschild metric is
\begin{equation}
ds^{2}=-(1-\frac{2m}{r})dt_{s}^{2}+(1-\frac{2m}{r})^{-1}dr^{2}+r^{2}
d \Omega ^{2}.\label{orimetric}
\end{equation}
By mode decomposition, the odd perturbation fields  $h_{03}$ and
$h_{13}$   can be written as
\begin{equation}
h_{03}=h_0(r)e^{-i\omega t_s}\left[\sin \theta
\frac{\partial}{\partial \theta}\right]P_{l}\left(\cos
\theta\right),\label{h03s}
\end{equation}
\begin{equation}
h_{13}=h_1(r)e^{-i\omega t_s}\left[\sin \theta
\frac{\partial}{\partial \theta}\right]P_{l}\left(\cos
\theta\right).\label{h13s}
\end{equation}
The dynamical perturbation equation, the Regge-Wheeler equation,
is
\begin{equation}
\frac{d^{2}Q}{dr^{*2}}+\left[k^{2}-V\right]Q=0,\label{ReggeWheeler}
\end{equation}
where the effective potential $V$ and the tortoise coordinate are
\begin{equation}
V=\left(1-\frac{2m}{r}\right)\left[\frac{l\left(l+1\right)}{r^{2}}-\frac{6m}{r^{3}}\right],
\end{equation}
\begin{equation}
r^{*}=r+2m\ln \left(\frac{r}{2m}-1\right)\label{tortoise}
\end{equation}
respectively. The quantity $Q$ is connected with the odd
perturbation fields $h_0$, $h_1$ by
\begin{equation}
h_{0}(r)=\frac{i}{k} \frac{d}{dr^{*}}\left(rQ\right)=\frac{i}{k}
\left[\left(1-\frac{2m}{r}\right)Q+r\frac{dQ}{dr^{*}}\right]\label{h0qrelation}
\end{equation}
and
\begin{equation}
h_{1}(r)=r\left(1-\frac{2m}{r}\right)^{-1}Q.\label{h1qrelation}
\end{equation}

Just as done in the reference \cite{vish}, we  first solve the
eq.(\ref{ReggeWheeler}) to get the odd perturbation  quantities
$h_{03}$, $h_{13}$ in the Schwarzschild metric coordinates for
simplification, then transform them to the Painlev\'{e} coordinate
frame to study the stable problem.

There are two independent solutions $f_1(r)$, $f_2(r)$ to the
eq.(\ref{ReggeWheeler}) with following asymptotic properties at
infinity $r^*\rightarrow \infty$:
\begin{equation}
f_{1}(r)\rightarrow A_1re^{-ikr^*},r^*\rightarrow
\infty,\label{f1}
\end{equation}
\begin{equation}
f_{2}(r)\rightarrow A_2re^{ikr^*}, r^*\rightarrow
\infty.\label{f2}
\end{equation}
$f_1(r)$, $f_2(r)$ corresponds to ingoing and out-going radiation
from infinity respectively. Similarly, there also are two
independent solutions $f_3(r)$, $f_4(r)$ to the
eq.(\ref{ReggeWheeler}), which have the  asymptotic forms
\begin{equation}
f_{3}(r)\rightarrow A_3re^{-ikr^*}, r^*\rightarrow -\infty
\label{f3}
\end{equation}
\begin{equation}
f_{4}(r)\rightarrow A_4re^{ikr^*} r^*\rightarrow
-\infty.\label{f4}
\end{equation}
at $r=2m$ or $r*\rightarrow -\infty$. $f_3(r)$, $f_4(r)$ are the
ingoing and out-going radiation from the black-hole horizon
respectively. The relations of the four solutions $f_1(r)$,
$f_2(r)$, $f_3(r)$, $f_4(r)$ are following:
\begin{equation}
f_{1}(r)=C_{13}f_3+C_{14}f_4,\label{f134}
\end{equation}
\begin{equation}
f_{2}(r)= C_{23}f_3+C_{24}f_4,\label{f234}
\end{equation}
or equivalently
\begin{equation}
f_{3}(r)=C_{31}f_1+C_{32}f_2,\label{f312}
\end{equation}
\begin{equation}
f_{2}(r)= C_{41}f_1+C_{42}f_2.\label{f412}
\end{equation}

The Schwarzschild metric is singular at $r=2m$, so, we could not
discuss the stable problem under this metric. To find the real
divergence at $ r=2m $ from the spurious one caused by the
improper choice of coordinate due to its ill-defined-ness at $
r=2m $ \cite{vish},  Vishveshwara studied the perturbation
quantities in the  reference frame regular at $r=2m$, that is, the
Kruskal reference frame.  Instead, the regular reference frame we
select is the Painlev\'{e} frame, which is stationary and  regular
at the horizon \cite{Krau}.

The Painlev\'{e} coordinate for the Schwarzschild black hole is
\begin{eqnarray}
ds^{2} = -\left(1-\frac{2m}{r}\right)dt_p^{2}
+2\sqrt{\frac{2m}{r}}drdt_p+dr^{2}+r^{2}d\Omega
^{2}.\label{painblackmetric}
\end{eqnarray}
The Painlev\'{e} metric (\ref{painblackmetric}) is obtained from
the Schwarzschild metric (\ref{orimetric}) by the
transformation\cite{Pari}-\cite{Krau}
\begin{equation}
t_{p}=t_{s}+\left[2\sqrt{2mr}+2m \ln
\frac{\sqrt{r}-\sqrt{2m}}{\sqrt{r}+\sqrt{2m}}\right],\label{paintran2}
\end{equation}
and is obviously regular, especially at the horizon $r=2m$.

In the usual Penrose diagram, part $I$ corresponds to our region($
r>2m $), parts $II$ and $II'$ are the black-hole($ r<2m $) and
white-hole($ r<2m $) respectively, and part $I'$ is another
region($ r>2m $) not communicating with our region. We define the
region $I$ connected with $II$ by the metric
(\ref{painblackmetric}) as the \textbf{\emph{black-hole-connected
region}}.

Similarly, the \textbf{\emph{white-hole-connected region}} is
defined as the region $I$ connected with $II'$ by the metric
\begin{eqnarray}
ds^{2} &=& -\left(1-\frac{2m}{r}\right)dt_p^{2}
-2\sqrt{\frac{2m}{r}}drdt_p+dr^{2}+r^{2}d\Omega ^{2} \nonumber\\
&=&
-dt_p^{2}+\left(dr-\sqrt{\frac{2m}{r}}dt_p\right)^{2}+r^{2}d\Omega
^{2}.\label{painwhitemetric}
\end{eqnarray}
The metric (\ref{painwhitemetric}) comes from the Schwarzschild
metric (\ref{orimetric}) by the
transformation\cite{Pari}-\cite{Krau}
\begin{equation}
t_{p}=t_{s}-\left[2\sqrt{2mr}+2m \ln
\frac{\sqrt{r}-\sqrt{2m}}{\sqrt{r}+\sqrt{2m}}\right].\label{paintran1}
\end{equation}

By equation (\ref{paintran2}) it is easy to get the perturbation
fields $[h^{p}_{ij}]$ in the metric (\ref{painblackmetric}) of
Painlev\'{e} coordinates:
\begin{equation}
h_{03}^{p}=h_{03}^{s}
\end{equation}
\begin{equation}
h_{13}^{p}=h_{13}^{s}-\sqrt{\frac{2m}{r}}\left(1-\frac{2m}{r}\right)^{-1}h_{03}^{s}.
\end{equation}
Using equations (\ref{h03s})-(\ref{h13s})) and
(\ref{h0qrelation})-(\ref{h1qrelation}), we could get,
\begin{equation}
h_{03}^{p}=\frac{i}{k}\left[\left(1-\frac{2m}{r}\right)Q
+r\frac{dQ}{dr^{*}}\right]e^{-ikt_s},\label{h0black}
\end{equation}
\begin{equation}
h_{13}^{p}=\{\frac{i}{k}\sqrt{\frac{2m}{r}}Q+r\left(1-\frac{2m}{r}\right)^{-1}
\left[-\frac{i}{k}\sqrt{\frac{2m}{r}}\frac{dQ}{dr^{*}}+Q\right]\}e^{-ikt_s}.\label{h1black}
\end{equation}

Now, we study the stable problem.

Corresponding to the in-going radiations $f_1$, $f_3$,  the
asymptotic forms of the fields $h_{03}$, $h_{13}$ are
\begin{eqnarray}
&&h^{p}_{03}(t_{p},r)\propto
\frac{1}{k}\left[(1-\frac{2m}{r})-ikr\right]
e^{-ik(t_s+r^{*})},\ r^* \rightarrow \pm \infty  \nonumber  \\
 &=& \frac{1}{k}\left[(1-\frac{2m}{r})-ikr\right]e^{-ikt_{p}}
 e^{ik\left[-r+2\sqrt{2mr}-4m\ln\left(1+\sqrt{\frac
r{2m}}\right)\right]}, \ r^* \rightarrow \pm \infty \label{h03-in}
\end{eqnarray}
and
\begin{eqnarray}
&&h^{p}_{13}(t_{p},r)  \propto \left[\frac{1}{k}\sqrt{\frac{2m}r}
+r\left(1-\frac{2m}{r}\right)^{-1}\left(\frac{1}{k}\sqrt{\frac{2m}{r}}(-k)+1\right)\right]
e^{-ik(t_s+r^{*})} ,  \ r^* \rightarrow \pm \infty \nonumber\\
 &&=
\left[\frac{1}{k}\sqrt{\frac{2m}{r}}+r(1+\sqrt{\frac{2m}{r}})^{-1}\right]
 e^{-ikt_{p}}
 e^{ik\left[-r+2\sqrt{2mr}-4m\ln\left(1+\sqrt{\frac
r{2m}}\right)\right]}, \ r^* \rightarrow \pm \infty \label{h13-in}
\end{eqnarray}
respectively. Similarly, corresponding to the out-going radiations
$f_2$, $f_4$, the asymptotic forms of the fields $h_{03}$,
$h_{13}$ are
\begin{eqnarray}
&&h^{p}_{03}(t_{p},r)\propto
\frac{1}{k}\left[(1-\frac{2m}{r})+ikr\right]
e^{-ik(t_s-r^{*})},\ r^* \rightarrow \pm \infty  \nonumber  \\
 &=& \frac{1}{k}\left[(1-\frac{2m}{r})+ikr\right]e^{-ikt_{p}}e^{2ikr^*}
 e^{ik\left[-r+2\sqrt{2mr}-4m\ln\left(1+\sqrt{\frac
r{2m}}\right)\right]}, \ r^* \rightarrow \pm \infty
\label{h03-out}
\end{eqnarray}
\begin{eqnarray}
&&h^{p}_{13}(t_{p},r)  \propto \left[\frac{1}{k}\sqrt{\frac{2m}r}
+r\left(1-\frac{2m}{r}\right)^{-1}\left(\frac{1}{k}\sqrt{\frac{2m}{r}}*k+1\right)\right]
e^{-ik(t_s-r^{*})} ,  \ r^* \rightarrow \pm \infty \nonumber\\
 &&=
\left[\frac{1}{k}\sqrt{\frac{2m}{r}}+r(1-\sqrt{\frac{2m}{r}})^{-1}\right]
 e^{-ikt_{p}}e^{2ikr^*}
 e^{ik\left[-r+2\sqrt{2mr}-4m\ln\left(1+\sqrt{\frac
r{2m}}\right)\right]} \ r^* \rightarrow \pm \infty \label{h13-out}
\end{eqnarray}
respectively.

First, to make $h_{03}^{p}(t_p,r)$, $h_{13}^{p}(t_p,r)$ regular
initially at infinity, we must have the following selections for
the frequency $k=k_1+ik_2$ with $k_1=\Re k$, $k_2=\Im k$:
$$
Q_{\infty}= \left\{
\begin{array}{cccc}
  f_1 & {\rm if} & k_2<0,\\
  f_2 &  {\rm if} & k_2>0,\\
B_1f_1+B_2f_2 &  {\rm if}& k_2=0
\end{array}
\right .
$$
Similarly, at the horizon, the regularity of the fields
$h_{03}^{p}(t_p,r)$, $h_{13}^{p}(t_p,r)$  requires
$$
Q_{2m}= \left\{
\begin{array}{cccc}
  f_3 & {\rm if} & k_2>-{\frac{1}{4m}},\\
  B_3f_3+B_4f_4 &  {\rm if} & k_2\leq -{\frac{1}{4m}}
\end{array}
\right .
$$

Therefore, if $k_2>0$, to make $h_{03}^{p}(t_p,r)$,
$h_{13}^{p}(t_p,r)$ well-behaved initially, we must select
$Q_{\infty}=f_2$ and $Q_{2m}=f_3$. From the equation (\ref{f234}),
we have
\begin{equation}
f_{2}(r)= C_{23}f_3+C_{24}f_4.\label{f234new}
\end{equation}
By the theorem of the quantum mechanics, it is easy to get
\[
C_{24}\neq 0.
\]
It is easy to see that there is a contradiction, so, $k_2>0$ is
impossible for the black-hole-connected region.

When $-\frac{1}{4m}<k_2<0$, we must select $Q_{\infty}=f_1$ and
$Q_{2m}=f_3$. This also a contradiction to the relation
(\ref{f134}). So, $-\frac{1}{4m}<k_2<0$ is impossible for
initially well-defined fields $h_{03}^{p}(t_p,r)$,
$h_{13}^{p}(t_p,r)$.

When $k_2\leq -\frac{1}{4m}$, then $Q_{\infty}=f_1$ and
$Q_{2m}=C_{13}f_3+C_{14}f_4$. This is a possibility to make the
fields $h_{03}^{p}(t_p,r)$, $h_{13}^{p}(t_p,r)$ well-behaved
initially, though the physical meaning is not too clear. In such
case, an ingoing radiation from infinity $f_1$ is scattered by the
black-hole's effective potential $V$, and the reflected radiation
$f_2$ is cancelled. There is the transmitted radiation $f_3$
falling into the black-hole. However, it is a strange thing that
there exists an outgoing radiation from the black-hole. What does
intrigue this kind of outgoing radiation from the black-hole?
Moreover, this kind radiation from the black-hole does transmit
the black-hole's effective potential $V$, and cancel the reflected
radiation from infinity. In this very case, the radiations all
diminish exponentially in time.

The last possibility is $k_2=0$. In this case, we have
$$
\left\{
\begin{array}{cccc}
  Q_{2m}&= &f_3   \\
  Q_{\infty }&= &C_{31}f_1+C_{32}f_2
\end{array}
\right .
$$
This is just what Vishveshwara have obtained in reference
\cite{vish1}. Its physical meaning is obvious: for an ingoing
radiation $f_1$ from infinity, some of it is reflected by the
hole, and the other transmits and falls into the black-hole ($f_3$
is the transmitted radiation, and $f_2$ is the reflected
radiation).

Summary: to make the fields $h_{03}^{p}(t_p,r)$,
$h_{13}^{p}(t_p,r)$ well-behaved initially in  the
\textbf{\emph{black-hole-connected region}}, we only have two
choices: (1) $\Im k <-\frac{1}{4m}$, or (2) $\Im k=0$. The
physical meaning corresponds to  the first choice is unclear yet,
the second have clear physical meaning. The conclusion is that the
\textbf{\emph{black-hole-connected region}} is stable according to
the definition of physicists', but is instable to the definition
of the mathematician' \cite{stew}.

Similar calculation could applied to the
\textbf{\emph{white-hole-connected region}}\cite{tian}. It is easy
to see that there exists initially well-behaved fields
$h_{03}^{p}(t_p,r)$, $h_{13}^{p}(t_p,r)$ with $\Re k=0$ and $Im
k>0$, so, the \textbf{\emph{white-hole-connected region}} is
unstable.

\section*{Acknowledgments}

We are supported in part by the National Science Foundation of
China under Grant No.10475013, No.10373003, No.10375087,
No.10375008, NKDRTC(2004CB318000) and the post-doctor foundation
of China.

\end{document}